\documentclass{elsarticle}
\usepackage{graphicx}
\usepackage{color}
\usepackage{amsmath}
\usepackage{amssymb}

\title{Quantum point contacts as heat engines}

\author{Sebastian Pilgram}
\address{Kantonsschule Frauenfeld, Ringstrasse 10, CH-8500 Frauenfeld, Switzerland}
\author{David S\'anchez\fnref{fn1}}
\ead{david.sanchez@uib.es}
\fntext[fn1]{Corresponding author: david.sanchez@uib.es}
\author{Rosa L\'opez}
\address{Instituto de F\'{\i}sica Interdisciplinar y Sistemas Complejos IFISC (UIB-CSIC), E-07122 Palma de Mallorca, Spain}

\begin{document}

\begin{abstract}
The efficiency of macroscopic heat engines is restricted by the second law of thermodynamics. They can reach at most the efficiency of a Carnot engine. In contrast, heat currents in mesoscopic heat engines show fluctuations. Thus, there is a small probability that a mesoscopic heat engine exceeds Carnot's maximum value during a short measurement time. We illustrate this effect using
a quantum point contact as a heat engine. When a temperature difference is applied to a quantum point contact, the system may be utilized as a source of electrical power under steady state conditions. We first discuss the optimal working point of such a heat engine that maximizes the generated electrical power and subsequently calculate the statistics for deviations of the efficiency from its most likely value. We find that deviations surpassing the Carnot limit are possible, but unlikely.
\end{abstract}

\maketitle

\noindent
M. B\"uttiker was among the first scientists to realize that measurements of current fluctuations deliver most valuable information about the internal structure of mesoscopic conductors \cite{Buttiker1992}. The measurement of shot noise \cite{Blanter2000} in a tunnel junction, for instance, may be used to determine the elementary charge of the charge carriers transferred through the circuit. Its measurement may as well serve to reveal the transmission probabilities of a multichannel mesoscopic point contact. 

The description of current fluctuations was later extended to full statistics of the charge transfer through a mesoscopic conductor \cite{Levitov1993}. From an experimental point of view, current fluctuations are probably the easiest to measure. Nevertheless, statistics for a number of other mesoscopic physical quantities have also been investigated: among them, combined charge-phase statistics in the superconducting state \cite{Belzig2001}, waiting time statistics of a closed volume \cite{Pilgram2003}, voltage statistics on a current biased point contact \cite{Kindermann2003}.

Recently, interest has shifted to energy transport through mesoscopic structures. The study of energy transport is partially motivated by the possibility to use small circuits to convert local temperature differences into voltages \cite{Sanchez2011}. Heat currents are subject to fluctuations as well \cite{Kindermann2004}. Such fluctuations have been theoretically studied in a number of situations~\cite{ave10,ser11,lim13,mos14,bat14}. Although the direct measurement of fluctuations in heat current is probably difficult, indirect consequences of energy fluctuations have been observed experimentally \cite{Steinbach1996,Henny1999,Oberholzer2001}.

In this paper, we apply the theory of heat current fluctuations to a question of rather conceptual than practical interest. We consider a mesoscopic heat engine that converts heat partially into electrical work. Since heat currents in mesoscopic devices fluctuate with time, any quantity derived from the heat currents will fluctuate as well. In particular, the efficiency of heat to work conversion will depend randomly on time and therefore may exceed the Carnot efficiency for a short time, not on average, but sometimes with a non-vanishing probability. It is the aim of this work to quantify this probability. We illustrate our discussion with a quantum point contact, a narrow constriction between two electrodes that shows quantized linear conductance.

\section{System and Formalism}
\begin{figure}[t]
\includegraphics[width=12cm]{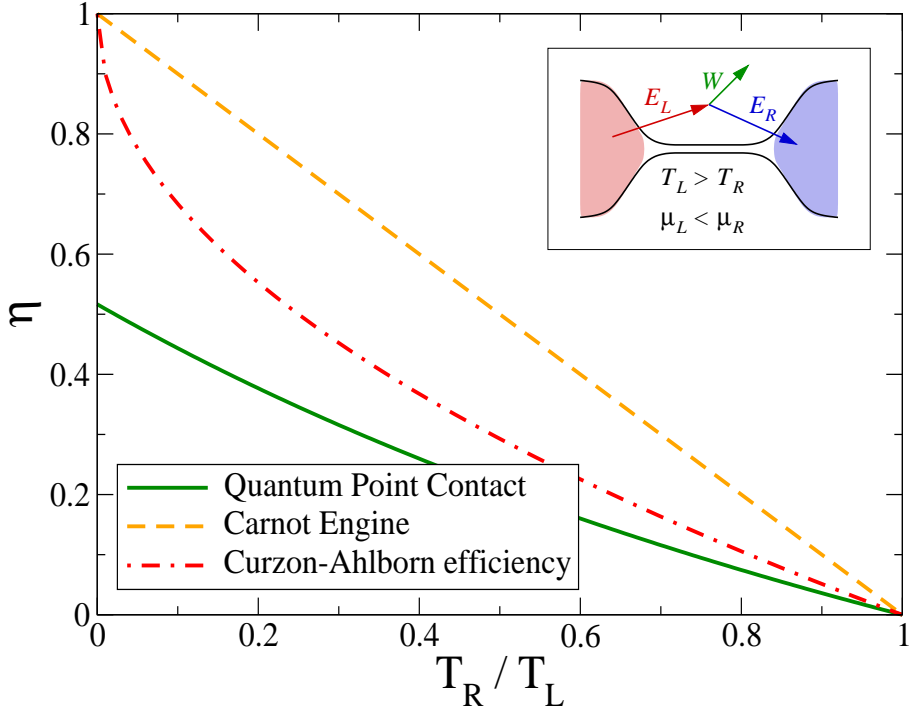}
\caption{ (Color online) {\bf Inset:} A quantum point contact connected to two reservoirs at different temperatures $T_{L}$, $T_R$ and chemical potentials $\mu_L$, $\mu_R$ may be used as a heat engine that converts heat $E_L$ partially into electrical work $W$. {\bf Main plot:} Efficiency $\eta=W/E_L$ versus the applied temperature ratio $T_R/T_L$ at the optimal working point, compared to the ideal efficiency of a Carnot process and the Curzon-Ahlborn efficiency at maximum power.}
\label{Point Contact}
\end{figure}

\noindent
Our system of interest is a mesoscopic point contact coupling left ($L$) and right ($R$) reservoirs that have in general different chemical potentials $\mu_L$, $\mu_R$ and different temperatures $T_L$, $T_R$. This is an experimentally relevant situation that has been the subject of recent works~\cite{mol90,mat12,mat14}. Quite generally, energy will flow from the left to the right reservoir when $T_L > T_R$. The energy flow is accompanied by a charge flow against a difference in the chemical potentials, $\mu_L < \mu_R$. This charge flow corresponds to the electrical work generated by the point contact. The inset of Fig.\ \ref{Point Contact} shows the setup and the sign conventions. The heat extracted from the left reservoir $E_L$ has a positive sign, the (smaller) heat evacuated into the right reservoir $E_R$ has a negative sign. We take the generated work $W$ as a positive quantity:
\begin{equation}
\label{Work Definition}
E_L + E_R = W >0\,.
\end{equation}
The efficiency of the heat engine is then defined as
\begin{equation}
\eta = \frac{W}{E_L} = \frac{E_L+E_R}{E_L}\,.
\end{equation}

We consider that both electrodes have well defined chemical potentials and temperatures. Hence, relaxation processes in the reservoirs are assumed to be fast compared to the time $\tau$ that our measurement takes. At the end of the experiment we record the amount of heat $E_L$ that is extracted from the hot reservoir and the amount of heat $E_R$ that is dumped into the cold reservoir. Due to thermal and quantum fluctuations, the whole transfer process is probabilistic and described by a probability distribution $P_\tau(E_L,E_R)$. Often it is more useful to use the corresponding cumulant generating function $S_\tau(i\xi_L,i\xi_R)$ instead of $P$. Both quantities are linked by Fourier transformation
\begin{equation}
P_\tau(E_L,E_R) \sim \int d\xi_L d\xi_R e^{-i(\xi_L E_L+\xi_R E_R) + S_\tau(i\xi_L,i\xi_R)}\,.
\end{equation}
The generating function describing the statistics of heat transfer in a general two-terminal conductor is given in Ref.~\cite{Kindermann2004}. Here, we adapt this generating function for our purpose:
\begin{align}
S_\tau(i\xi_L,i\xi_R) & = \frac{\tau}{\pi} \int d\varepsilon \ln \Bigl\{ 1 + 
\Gamma f_L(1-f_R) \left(e^{i\xi_L(\varepsilon-\mu_L)-i\xi_R(\varepsilon-\mu_R)}-1\right)\nonumber\\
& + \Gamma (1-f_L)f_R \left(e^{-i\xi_L(\varepsilon-\mu_L)+i\xi_R(\varepsilon-\mu_R)}-1\right)
\Bigr\}\,.\label{Generating Function}
\end{align}
This expression applies to a spin-degenerate single-channel point contact with transparency $\Gamma$. It also contains
the leads' Fermi occupation factors $f_{L,R} = 1/[1+e^{(\varepsilon - \mu_{L,R})/T_{L,R}}]$. For convenience, we hereafter choose units such that the Planck constant, the unit charge and the Boltzmann constant are equal to one. Any cumulant of the distribution $P_\tau(E_L,E_R)$ can be obtained from the generating function by taking partial derivatives
\begin{equation}
\langle \langle (E_L)^m (E_R)^n \rangle \rangle = \left.\frac{\partial^{m+n}S}{(\partial i\xi_L)^m (\partial i\xi_R)^n}\right|
_{\xi_L=\xi_R=0}\,.
\end{equation}

Thermoelectric effects necessarily require that the transmission probability $\Gamma$ of the quantum point contact depends on energy $\varepsilon$. A basic and convenient model for this dependence was proposed by M.\ B\"uttiker \cite{Buttiker1990}. If the potential barrier creating the point contact is a saddle, the transmission probability of one single transmission channel reads
\begin{equation}
\Gamma(\varepsilon) = \frac{1}{1+e^{-(\varepsilon - \varepsilon_0)/\omega_z}}\,,
\end{equation}
where $\varepsilon_0$ is the potential at the saddle and $\omega_z$ gives the energy width of the transition region.
In this work we will use a simplified version. We assume a very long point contact, $\omega_z \rightarrow \infty$, and choose the energy scale such that $\varepsilon_0=0$. The
transmission probability then jumps sharply from zero to one
\begin{equation}
\label{Transmission Model}
\Gamma(\varepsilon) = \left\{
\begin{array}{cc}
0 & \varepsilon < 0\\
1 & \varepsilon > 0
\end{array}
\right.\,,
\end{equation}
when energy surpasses the threshold set by $\varepsilon_0$.

\section{Optimal Working Point}\label{sec_opt}
\noindent
For a given set of temperatures $T_L$ and $T_R$ we may optimize the chemical potentials $\mu_L$ and $\mu_R$ such that production of work is maximized on average. We call this optimized situation the working point of the point contact.
The result is equivalent to that of the efficiency at maximum power
which has been analyzed in detail in Ref.~\cite{whi14}
in the context of scattering theory of quantum transport.
From Eq.~\eqref{Work Definition} we have
\begin{equation}
\label{Working Point}
\langle\langle W  \rangle\rangle = \langle\langle E_R\rangle\rangle + \langle\langle E_L\rangle\rangle,\qquad
\frac{\partial \langle\langle W \rangle\rangle}{\partial \mu_{L,R}} = 0\,,
\end{equation}
with the heat extracted from the heat reservoir given by
\begin{equation}
\langle\langle E_L \rangle\rangle \sim \tau \int_0^\infty d\varepsilon \left(f_L-f_R \right) \left(\varepsilon-\mu_L\right)\,,
\end{equation}
and the generated work obeying the Joule expression
\begin{equation}
\langle\langle W \rangle\rangle \sim \tau \left(\mu_R-\mu_L\right) \int_0^\infty d\varepsilon \left(f_L-f_R \right)\,,
\end{equation}
for the nonzero voltage difference $\mu_R-\mu_L$.
Combining the derivatives with respect to both chemical potentials we find 
\begin{equation}\label{mulmur}
\frac{\mu_L}{T_L} = \frac{\mu_R}{T_R}\,.
\end{equation}
Since $T_L>T_R$ and $\mu_L<\mu_R$ (charge current must flow against the potential from the hot to the cold reservoir), it follows immediately that both chemical potentials have to be negative (we set the Fermi energy $E_F=0$). Substituting Eq.~\eqref{mulmur} into $\langle\langle W \rangle\rangle$ and calculating the energy integral yields the intermediate result
\begin{equation}
\langle \langle W \rangle \rangle \sim -\tau \mu_L \frac{(T_L-T_R)^2}{T_L}
\ln\left( 1 + e^{\mu_L/T_L} \right)\,,
\end{equation}
which still needs to be maximized with regard to $\mu_L$. We define $x=e^{\mu_L/T_L}$ and obtain a symmetric transcendental equation and its numerical solution
\begin{equation}
\label{Optimal Working Point}
(1+x)\ln(1+x)=-x\ln(x), \qquad x \simeq 0.318, \qquad \mu_{L,R} \simeq -1.14T_{L,R}\,.
\end{equation}
Interestingly, both potentials lie below the subband energy $\varepsilon_0$ and therefore the Fermi functions
could be approximated as Boltzmann factors. This fact demonstrates that the operational principle of our device is based on a classical distribution of electrons.

For a compact representation of the results, we introduce the numerical constants
\begin{equation}
y = -\ln(x)\ln(1+x) \simeq 0.316, \quad z = \int_{x^{-1}}^\infty \frac{dt \ln t}{t(1+t)} \simeq 0.612, \quad
\frac{y}{z} \simeq 0.517.
\end{equation}
The mean work and extracted heat for the working point become
\begin{align}
\langle\langle W \rangle\rangle_w &= y \frac{\tau}{\pi}\left(T_L-T_R\right)^2 \,,\\
\langle\langle E_L \rangle\rangle_w &= z \frac{\tau}{\pi} \left(T_L-T_R\right)\left\{T_L+\left(1-\frac{y}{z}\right)T_R\right\}\,.
\end{align}
With these results we are ready to calculate the mean efficiency at maximum power extraction
\begin{equation}\label{eta}
\langle\langle \eta \rangle\rangle_w = \frac{y}{z} \frac{1-\frac{T_R}{T_L}}{1+\left(1-\frac{y}{z}\right)\frac{T_R}{T_L}}
\simeq
0.517 \frac{\eta_C}{1.483-0.483\eta_C}\,.
\end{equation}
in terms of the Carnot efficiency $\eta_C=1-T_R/T_L$. Equation~\eqref{eta} has been derived in Ref.~\cite{whi14} using nonlinear scattering theory.
The mean efficiency reaches at most $50$\% of the Carnot efficiency. The efficiency of heat engines operated at maximum power is additionally limited by the Curzon-Ahlborn efficiency~\cite{ca,bro05}, $\eta_{CA}=1-\sqrt{T_R/T_L}$, which sets a stronger bound than the Carnot efficiency. An expansion of Eq.~\eqref{eta},
$\langle\langle \eta \rangle\rangle_w =0.349 \eta_C+0.114 \eta_C^2$, shows that
our result falls below the Curzon-Ahlborn efficiency and approaches $\eta_{CA}$
only near equilibrium.

For comparison, in Fig.\ \ref{Point Contact}
we show $\langle\langle \eta \rangle\rangle_w$ (full line), $\eta_C$ (dashed line) and $\eta_{CA}$ (dot-dashed line)
as a function of the temperature ratio. In the strong equilibrium case, $\langle\langle \eta \rangle\rangle_w$ reaches roughly
half the Carnot efficiency. We emphasize at this stage that we are considering an idealized model. The obtained efficiency serves as a reference point for the upcoming calculation. The experimentally measurable efficiency of a realistic constriction would be probably much lower, although quantum point contacts perform rather well as compared with other low-dimensional systems~\cite{nak10}.

\section{Distribution of the Efficiency}

\noindent
We have thus discussed the mean efficiency of average heat to average work conversion. However, the efficiency of a mesoscopic
heat engine in general fluctuates because both work and heat fluctuate over time. Therefore, efficiencies larger than
the Carnot limit are in principle possible~\cite{ver14}. To see this, we calculate the efficiency distribution by integrating the complete distribution $P_\tau(E_L,E_R)$ with respect to one variable
\begin{equation}
P_\tau(\eta) \sim \int dE_L P_\tau(E_L,-(1-\eta)E_L)\,.
\end{equation}

After representing the complete distribution by its generating function, Eq.~(\ref{Generating Function}), the energy integration can be carried out. This step fixes one auxiliary field $\xi_L = (1-\eta)\xi_R$ and leaves us with a single integration
\begin{equation}
P_\tau(\eta) \sim \int d\xi_R e^{S_\tau(i(1-\eta)\xi_R,i\xi_R)}\,.
\end{equation}
We evaluate the last integral to exponential accuracy by deforming the integration path in the complex plane and by using a saddle point approximation, i.e., replacing the integral by its value at the saddle point $i\xi_R$,
\begin{equation}
\frac{\ln P_\tau(\eta)}{\tau T_L /\pi} \simeq \mbox{min}_{i\xi_R} \bigl[ S_\tau((1-\eta)i\xi_R,i\xi_R) \bigr]\,.
\end{equation}

\begin{figure}[t]
\includegraphics[width=11cm]{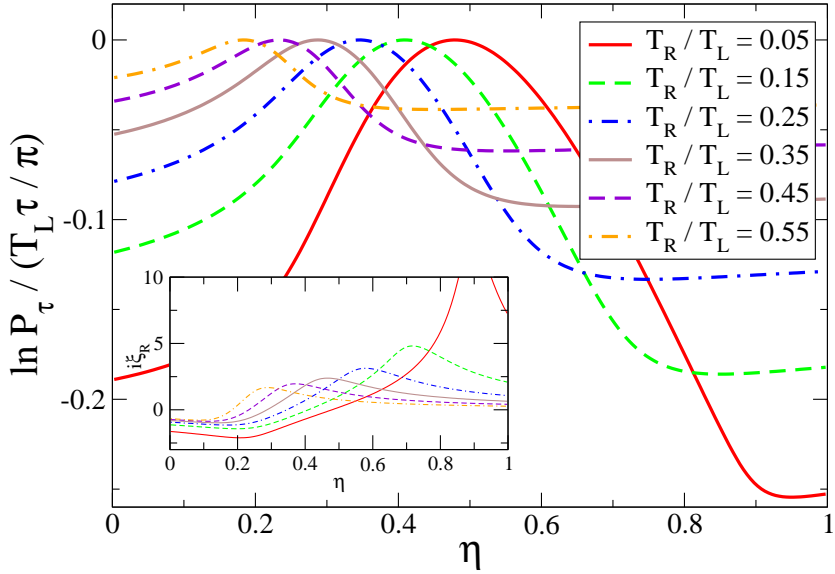}
\caption{(Color online) {\bf Main plot:} Distribution $P_\tau$ of the efficiency $\eta = W / E_L$ on a logarithmic scale. The Carnot efficiency $1-T_R/T_L$ can be exceeded with an exponentially small probability. {\bf Inset:} Saddle point solution $\xi_R$ used to calculate the distributions shown in the main plot.}
\label{Efficiency Distribution}
\end{figure}

The minimization is carried out numerically. Our results for the distribution $P_\tau(\eta)$ are shown in Fig.\ \ref{Efficiency Distribution} for different temperature ratios $T_R/T_L$. The chemical potentials are taken such that the point contact is operated at its optimal working point as discussed in Sec.~\ref{sec_opt}. The inset of the figure displays the saddle point solutions $i\xi_R(\eta)$ corresponding to the different distributions. 

We observe in Fig.~\ref{Efficiency Distribution} that the probability peaks at the mean efficiency $\langle\langle \eta \rangle\rangle_w$. At the larger corresponding Carnot efficiency, it shows a weak minimum~\cite{ver14}.
This minimum is most pronounced for strong nonequilibrium ($T_R/T_L=0.05$).
The Carnot limit is attained in the exponential tails of the distribution. It is therefore very unlikely that our device
beats the second law. A read-off example illustrates that the expected effect is indeed weak. Let us consider a fictitious measurement lasting $10^{-8}$~s and taking place at temperatures $T_L=1000$~mK and $T_R=450$~mK (purple dashed curve in Fig.~\ref{Efficiency Distribution}). For these values, the efficiency at maximum power given by Eq.~\eqref{eta}
is $\langle\langle \eta \rangle\rangle_w=0.23$ (the mean efficiency at this temperature ratio) whereas the Carnot efficiency
is $\eta_{C}=0.55$. Now, taking into account the units of the Boltzmann constant $k_B$ and Planck's constant $\hbar$, the probability $P_\tau(1)d\eta$ of an efficiency of 100\% (larger than the Carnot value) is found to be roughly $10^{11}$ times smaller than the maximal probability $P_\tau(0.23)d\eta$ corresponding to $\langle\langle \eta \rangle\rangle_w=0.23$.
It would thus take a total measurement time of $1000$~s to measure such a second-law violating event once.

\section{Conclusions}

To sum up, we have considered a mesoscopic heat engine (a quantum point contact) coupled to hot and cold
reservoirs and operating in a nonequilibrium steady state. We have determined the efficiency that maximizes the
work extraction in terms of the reservoir temperatures. Our result is based on a transmission function that depends
on energy and is valid for noninteracting electrons. M.\ B\"uttiker soon recognized that in the nonlinear regime
of transport electron-electron interactions should be taken into account within a gauge-invariant current-conserving
scattering theory~\cite{nonlinear}. When the potential landscape is self-consistently calculated, the transmision
probability becomes a function of the applied temperature and voltage differences~\cite{san13,mea13,whi13}.
This effect has been neglected here, which is a valid approximation for quantum channels strongly coupled
to a nearby gate~\cite{but03}. Nevertheless, we leave the question open as to what extend our results will be modified
in the presence of nonlinearities and interactions.

Furthermore, we have discussed departures of the second law of thermodynamics due to fluctuations of the efficiency
that exceed the Carnot limit. For a quantum point contact, we have determined the full distribution of the efficiency
and have quantified the probability that a given measurement deviates from the mean efficiency value. Our results
are thus relevant for a careful assessment of the heat-to-work conversion properties 
of quantum coherent heat engines.

\section*{Acknowledgments}

This work has been supported by MINECO under Grant No.\ FIS2014-52564.


\end{document}